\documentclass[prd,a4paper,onecolumn]{revtex4-2}
\usepackage{graphicx}
\usepackage{amsfonts}
\usepackage{amssymb}
\usepackage{amsmath} 
\usepackage{empheq}

\newcommand{\mt}{\tilde{f}}
\newcommand{\mtp}{\tilde{P}}
\newcommand{\be}{\begin{equation}}
\newcommand{\ee}{\end{equation}} 
\newcommand{\ben}{\begin{displaymath}}
\newcommand{\een}{\end{displaymath}} 
\newcommand{\ba}{\begin{eqnarray}}
\newcommand{\ea}{\end{eqnarray}} 
\newcommand{\ban}{\begin{eqnarray*}}
\newcommand{\ean}{\end{eqnarray*}} 

\newcommand{\de}{\partial}

\newcommand{\der}{-\frac{\pi}{\alpha}\frac{\de}{\de \log\epsilon}}

	\newcommand{\ff}[2]{f_{ 
  \scriptscriptstyle #1 \, \scriptscriptstyle #2}}
	\newcommand{\Ff}[2]{{\tilde{ f}}_{ 
  \scriptscriptstyle #1 \, \scriptscriptstyle #2}}

	\newcommand{\eps}{\varepsilon}
	\newcommand{\lzp}{L_0^+}\newcommand{\lzm}{L_0^-}\newcommand{\lup}{L_1^+}
	\newcommand{\Lzp}{L_0^+}
\newcommand{\Lzm}{L_0^-}\newcommand{\Lup}{L_1^+}\newcommand{\Lum}{L_1^-}
\newcommand{\Gdp}{G_2^+}\newcommand{\Gzp}{G_0^+}
\newcommand{\Gum}{G_1^-}
\newcommand{\kt}{\mbox{$k$}_{\perp }}

\usepackage[usenames]{color} 
\usepackage{ulem}

\begin{document}
\title{
Electroweak evolution equations and isospin conservation
}

%\date{\today}
\author{Paolo Ciafaloni$^{a}$}
\email{paolo.ciafaloni@le.infn.it}
\author{Giampaolo Co'$^{a}$}
\email{gpco@le.infn.it}
\author{Dimitri  Colferai$^{b}$}
\email{colferai@fi.infn.it}
\author{Denis  Comelli$^{c}$}
\email{comelli@fe.infn.it}
\affiliation{$^a$INFN e Università del Salento - Lecce, Italy}
%\affiliation{$^b$INFN e Università del Salento - Lecce, Italy}
\affiliation{$^b$INFN e Università di Firenze - Firenze, Italy}
\affiliation{$^c$INFN Sezione di Ferrara, Italy}

%%%%%%%%%%%%%%%%%%%%%%%%%%%%%%%%%%%%%%

\begin{abstract}
\noindent 
In processes taking place at energies much higher than the weak scale, electroweak corrections can be taken into account by using  electroweak evolution equations, that are
analogous to the DGLAP equations in QCD.
We show that weak isospin conservation   in these equations
imposes to modify the expressions of the splitting functions commonly used in the literature. These modifications 
have a profound impact on the parton distribution functions.
\end{abstract}
%%%%%%%%%%%%%%%%%%%%%%%%%%%%%%%%%%%%%%
\maketitle
%%%%%%%%%%%%%%%%%%%%%%%%%%%%%%%%%%%%%%

\section{Introduction}\label{sec:Intro}
Electroweak evolution equations, which are analogous to the
Dokshitzer, Gribov, Lipatov, Altarelli, Parisi (DGLAP) 
equations in Quantum Chromodynamics (QCD) \cite{DGLAP}, 
are of primary importance when the energies of the processes considered are much larger than those of the weak scale.
Indeed, at center-of-mass energies $Q$ much higher than the electroweak
(EW) symmetry breaking scale $M\sim$ 100 GeV, radiative EW corrections grow like
$\log^2 (Q/M)$ \cite{Ciafaloni:1998xg,verza}. One loop corrections reach the 30\% level at the TeV scale, and, for this reason, keeping the perturbative series under control is challenging \cite{NLL}, and will be particularly important for next generation of very high energy colliders \cite{collider}. Moreover such EW corrections are present even for fully inclusive quantities \cite{Ciafaloni:2000df} in contrast with QCD where large cancellations between real and virtual corrections take place, and are therefore ubiquitous whenever  the initial state is charged under SU(2). With the purpose of addressing these issues, EW evolution equations have been developed in \cite{Ciafaloni:2001mu,Ciafaloni:2005fm}. 
These  new equations
allow the resummation of all the terms ${\cal O} [\alpha\log^{2} (Q / M) ]^n$
 of infrared/collinear origin,  and of the ${\cal O}[\alpha\log (Q / M)]^n$ 
 terms of collinear 
 origin (actually,  terms of order $\alpha^n \log^k(Q/M)$ with $n\le k\le 2n$ are resummed).
The EW evolution equations are integro-differential equations involving the so called 
splitting functions, that are the kernels of the equations themselves. 

The main point of the present work is to advocate the use of splitting functions that differ from those used until now in the literature since they include a cutoff near $z=0$, $z$ being the momentum fraction. For instance, in the case of the kernel $P_{gf}^R$ that describes the splitting of a fermion $f$ into a gauge boson $g$ and a final state fermion, we have that
\be
\frac{1+(1-z)^2}{z}\to \frac{1+(1-z)^2}{z}\theta\left(z- \frac{\mu}{Q} \right)
\;\;,
\ee
where in the left hand side there is the standard expression of the splitting function 
and in the right hand side the one we propose here. 
The variable  $\mu$ indicates
the soft sliding scale with respect to which the functions are evolved (see section \ref{sec2}). As we explain in  section \ref{sec3}, the need to modify    the  splitting functions arises on one hand from   the probabilistic interpretation of the Parton Distribution Functions (PDFs), and on the other hand from  quantum numbers conservation (symmetries of the theory).

 The introduction of the modified splitting functions produces sizeable effects on the PDFs, as we show in section \ref{sec4}. Indeed, the distribution functions we obtain differ significantly from the standard ones, not only in the region close to $z=0$ but also for $z$ of order 1. In section \ref{sec5} we discuss a point which has been overlooked in the literature so far: the equivalence between the ultraviolet
(UV) evolution equations, that are evolved with respect to a hard scale $q$, and the 
infrared (IR) evolution equations, where the running scale is a soft one $\mu$. We show that the two approaches are indeed equivalent, but they produce the same PDFs only with an appropriate choice of cutoffs.

%%%%%%%%%%%%%%%%%%%%%%%%%%%
\section{\label{sec2}The Electroweak Evolution Equations}

\begin{figure}[htb]
   \centering
    \includegraphics[scale=0.6] {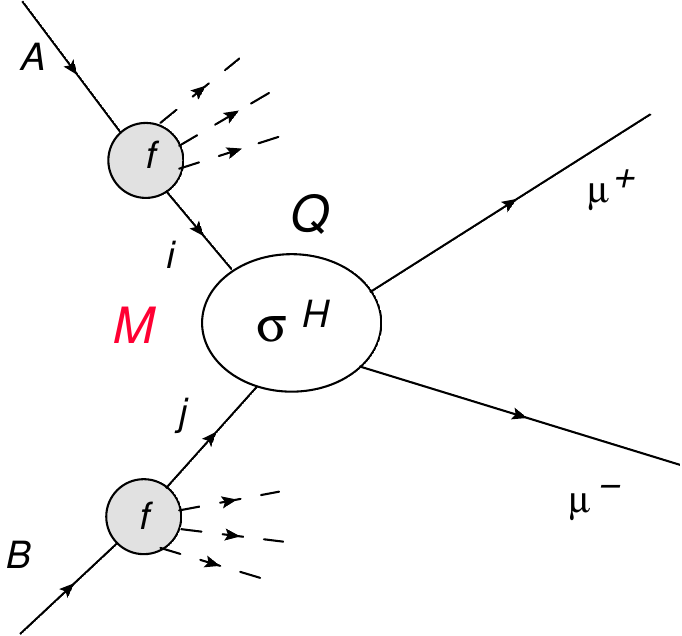}
    \caption{The overall cross section $\sigma(AB\to\mu^+\mu^-+X)$ can be written as a convolution of PDFs $f_{ij}s$ and a Hard partonic cross section (see text)
 }  
\label{fig:factorization}  
\end{figure}
In this work we consider a overall cross section $\sigma(AB\to\mu^+\mu^-+X)$ involving initial states provided by the collider $A,B$ (that can be leptonic and/or hadronic) and a final state with a tagged $\mu^+\mu^-$ state and completely inclusive over emitted radiation X. 
We can write:
$$
\sigma(AB\to\mu^+\mu^-+X)=\sum_{i,j}\int dx_i dx_j \, f_{iA}(x_i,{ M})\;\sigma_{ij}^H(x_jx_iQ^2)\;f_{jB}(x_j,{ M}) ,
$$
$\sigma^H$ being the hard (partonic) cross serction. Note that X includes, besides the customary QED and QCD radiation, also EW gauge bosons with their decay products.  ElectroWeak Evolution Equations (EWEEs)  describe the scale dependence of the  PDFs
 $f_{ij}$, representing the distribution of parton $i$ inside parton $j$. 
These functions depend on the momentum fraction $x$ 
of the daughter particle and on $\eps= \mu / Q$, where $\mu$ 
is a running scale with the physical meaning of lower cutoff for 
the daughter's transverse momentum.
The PDFs $f_{ij}(x,\mu/Q)$ are evaluated by solving the evolution equations 
starting from the initial condition at $\mu=Q$ and letting them evolve until 
$\mu$ reaches the weak scale value $M$. At this point, 
we obtain the final result $f_{ij}(x, M / Q)$.

In this work, we use a subset of the Standard Model (SM) equations, 
specifically we consider only left leptons and we set the U(1) coupling constant
$g'$ equal to zero. As a consequence, we have to deal
with only 7 different partons, that are gauge eigenstates: 
the left fermions electron $e$ and neutrino $\nu$, 
the corresponding antifermions $\bar{e}, \bar{\nu}$ 
and the transverse gauge bosons $W_-,W_3,W_+$. 
This limitation makes
our results not suitable to be directly compared with experimental data.
However, the purpose of our work consists in 
clearly showing that the splitting functions in the EW evolution equations
need to be modified and  that these modifications have a relevant effect on the PDFs.

We found convenient to use a basis of definite isospin and CP quantum numbers. We label  $L_T^{CP}$ the left fermion eigenstate with isospin $T$ and definite $CP$; in analogy, we indicate the transverse gauge boson states as $G_T^{CP}$.
We adopt the classification of states defined in \cite{Ciafaloni:2005fm}, with respect to which we use a slightly different normalization:
\be\label{translateL0}
f_{*\, \Lzp}=\frac{f_{*\,\nu}+f_{*\,e}+f_{*\,\bar{\nu}}+f_{*\,\bar{e}}}{2},\quad
f_{*\, \Lzm}=\frac{f_{*\,\nu}+f_{*\,e}-f_{*\,\bar{\nu}}-f_{*\,\bar{e}}}{2}
\;\;,
\ee
\be\label{translateL1}
f_{*\, \Lup}=\frac{f_{*\,\nu}-f_{*\,e}+f_{*\,\bar{\nu}}-f_{*\,\bar{e}}}{2},\quad
f_{*\, \Lum}=\frac{f_{*\,\nu}-f_{*\,e}-f_{*\,\bar{\nu}}+f_{*\,\bar{e}}}{2}
\;\;,
\ee
\be\label{translateL2}
f_{*\, \Gzp}=\frac{f_{*\,W_+}+f_{*\,W_3}+f_{*\,W_-}}{\sqrt{3}},\quad
f_{*\, \Gum}=\frac{f_{*\,W_+}-f_{*\,W_-}}{\sqrt{2}},\quad
f_{*\, \Gdp}=\frac{f_{*\,W_+}-2f_{*\,W_3}+f_{*\,W_-}}{\sqrt{6}}
\;\;,
\ee
where the asterisk stands for a generic index and similar expressions hold when we keep the right index fixed. We use capital letters for $T,CP$ eigenstates:
$A=(\Lzp,\Lzm,\Lup,\Lum,\Gzp,\Gum,\Gdp)$ and small letters for gauge eigenstates
$i=(\nu,e,\bar{\nu},\bar{e},W_+,W_3,W_-)$. 
The translation from one basis to the other is then given by a mixed indices unitary matrix $U_{Ai}$ such that
$f_{AB}=U_{Ai}U_{Bj}f_{ij}$ and
\be\label{transformer}
U=\begin{pmatrix}
\frac{1}{2} & \frac{1}{2} & \frac{1}{2} & \frac{1}{2} & 0 & 0& 0\\
\frac{1}{2} & \frac{1}{2} & -\frac{1}{2} & -\frac{1}{2} & 0 & 0& 0\\
\frac{1}{2} & -\frac{1}{2} & \frac{1}{2} & -\frac{1}{2} & 0 & 0& 0\\
\frac{1}{2} & -\frac{1}{2} & -\frac{1}{2} & \frac{1}{2} & 0 & 0& 0\\
0 & 0 & 0 & 0 & \frac{1}{\sqrt{3}} & \frac{1}{\sqrt{3}} & \frac{1}{\sqrt{3}}\\
0 & 0 & 0 & 0 & \frac{1}{\sqrt{2}} & 0 & -\frac{1}{\sqrt{2}}\\
0 & 0 & 0 & 0 & \frac{1}{\sqrt{6}} & -\frac{2}{\sqrt{6}} & \frac{1}{\sqrt{6}}
\end{pmatrix}
\;\;.
\ee

The matrix $f$ is a $7 \times 7$ square matrix, having, therefore 49 different matrix elements. 
However, working in the total isospin $t$-channel, due to isospin and CP conservation, 
the majority of these elements 
vanish. In fact, isospin $T$ and  $CP$ conservation implies
\be\label{ftcp}
f_{A_{T_1}^{CP_1}\, B_{T_2}^{CP_2}}(x,\,\eps)=\delta_{T_1\,T_2}\; \delta_{CP_1\,CP_2}\;\;
f_{A_{T_1}^{CP_1}\, B_{T_1}^{CP_1}}(x,\,\eps),\qquad \forall \;A,B=L,\,G
\ee
for a total of 11 independent PDFs that can be grouped in the following combinations:
\ba\label{totT}
&&(T=0,CP=+):\begin{array}{|cc|}
 f_{L_{0}^{+}\,L_{0}^{+}}  & f_{L_{0}^{+}\,G_{0}^{+}} \\
f_{G_{0}^{+}\,L_{0}^{+}} & f_{G_{0}^{+}\,G_{0}^{+}} 
\end{array}\;,
\quad  (T=1,CP=-):\begin{array}{|cc|}
 f_{L_{1}^{-}\,L_{1}^{-}}  & f_{L_{1}^{-}\,G_{1}^{-}} \\
 f_{G_{1}^{-}\,L_{1}^{-}} & f_{G_{1}^{-}\,G_{1}^{-}} 
\end{array}\;,
\\&&\nonumber
\\&&\nonumber
(T=0,CP=-):f_{L_{0}^{-}\,L_{0}^{-}},\quad
(T=1,CP=+): f_{L_{1}^{+}\,L_{1}^{+}} \;,
 \quad 
(T=2,CP=+):f_{G_{2}^{+}\,G_{2}^{+}}
\;\;.
\ea

By considering the various reaction channels, for the IR evolution equations
we have the following kernels depending on the variable $\eps$ defined above:
\be\label{split1}
P_{ff}^V=-\delta(1-z)\left(\log\frac{1}{\eps^2}-\frac{3}{2}\right);
\quad
P_{ff}^R=\frac{1+z^2}{1-z}\theta(1-\eps-z)
\;\;,
\ee
\be\label{split2}
P_{gg}^V=-\delta(1-z)\left(\log\frac{1}{\eps^2}-\frac{5}{3}\right);
\;\;
P^R_{gg}=2\left(z\;(1-z)+\frac{z}{1-z}+\frac{1-z}{z}
\right)\;\big[\theta(z-\eps)\big]\;\theta(1-\eps-z)
\;\;,
\ee
\be\label{split3}
P_{gf}^R=\frac{1+(1-z)^2}{z}\;\big[\theta(z-\eps)\big];
\quad
P_{fg}^R=z^2+(1-z)^2
\;\;,
\ee
 where with the upper index $R$ (for real) or $V$ (for virtual)
 we denote the origin of the corresponding contributions.
 The terms between square brackets
 in Eqs.~(\ref{split2}, \ref{split3}) are not present in the usual expressions of the kernels,
and produce  the threshold effect proportional to $\theta(z-\eps)$.
In the next section we will discuss the origin of these terms.

Schematically, the equations in the two different basis can be written as:
\be\label{sola}
\der f_{ij}(x,\eps)=[f_{ik}\otimes P^G_{kj}](x,\eps)\;\; , \qquad  \qquad
\der f_{AB}(x,\eps)=[f_{AC}\otimes P^I_{CB}](x,\eps)
\;\;,
\ee
where we have defined the convolution:
\be
\label{convolution}
[f \otimes P ](x,\eps)=\int_x^1\frac{dz}{z}f(z,\eps)P\left(\frac{x}{z},\eps\right)
\;.
\ee

In the above equations, $P^G$ is a $7 \times 7$ matrix with all elements different from zero. 
 Also  $P^I$ is a $7 \times 7$ matrix, but it is block-diagonal: as can be seen from Eq.~(\ref{totT})
it is a $2 \times 2$ submatrix in the $T=0,CP=+1$ subspace and in the $T=1,CP=-1$ subspace, while it is a 
$1 \times 1$ submatrix in the $T=1,CP=+1;T=0,CP=-1;T=2,CP=+1$ subspaces. 
The evolution equations have therefore a much simpler form in the isospin basis; we now write such equations. 
We use the values of $P^I$ taken from \cite{Ciafaloni:2005fm} and obtain:
\ba\label{eqini}
 \der \ff{\lzm}{\lzm}(x,\eps)&=&
\frac{3}{4}  [\ff{\lzm}{\lzm}\otimes(P_{ff}^R+P^V_{ff})](x,\eps)
\;\;,
\\
\der \ff{\lup}{\lup}(x,\eps) &=&
[\ff{\lup}{\lup}\otimes P^V_{ff}](x,\eps)-
\frac{1}{4}\ff{\lup}{\lup}\otimes(P_{ff}^R+P^V_{ff})
\label{eqT1p}
\;\;,
\\
\der \ff{\Gdp}{\Gdp}(x,\eps)&=&
3 [\ff{\Gdp}{\Gdp}\otimes P^V_{gg}](x,\eps)- 
 [\ff{\Gdp}{\Gdp}\otimes(P_{gg}^R+P^V_{gg})](x,\eps)
 \;\;.
\ea
In the $0^+$ channel we have 2 sets of $2 \times 2$ systems, corresponding to the possible values $A=L_0^+$ and  $A=G_0^+$:

\begin{equation}
\left\{ \begin{aligned} 
\der \ff{A}{\lzp}(x,\eps)&=
\frac{3}{4}[\ff{A}{\lzp}\otimes(P_{ff}^R+P^V_{ff})](x,\eps)+
\frac{\sqrt{3}}{2}[\ff{A}{\Gzp}\otimes
P^R_{gf}](x,\eps)
\;\;,
\\\label{eqnonso}
\der \ff{A}{\Gzp}(x,\eps)&=\frac{\sqrt{3}}{2}
[\ff{A}{\Lzp}\otimes
P^R_{fg}](x,\eps)+
2 [\ff{A}{\Gzp}\otimes(P_{gg}^R+P^V_{gg})](x,\eps)
\;\;.
 \end{aligned} \right.
\end{equation}
The same happens in the $1^-$ channel, with $A=L_1^-$ and  $A=G_1^-$: 
\begin{equation}
\left\{ \begin{aligned} 
\der \ff{A}{L_{1}^-}(x,\eps)&=
[\ff{A}{\Lum}\otimes P^V_{ff}](x,\eps)-
\frac{1}{4}[\ff{A}{L_1^-}\otimes(P_{ff}^R+P^V_{ff})](x,\eps)+
\frac{1}{\sqrt{2}}[\ff{A}{\Gum}\otimes
P^R_{gf}](x,\eps)
\;\;,
\\\label{eq1m}
\der \ff{A}{\Gum}(x,\eps)&=\frac{1}{\sqrt{2}}[\ff{A}{\Lum}\otimes P^R_{fg}](x,\eps)
 +[\ff{A}{\Gum}\otimes P^V_{gg}](x,\eps)+[\ff{A}{\Gum}\otimes(P_{gg}^R+P^V_{gg})](x,\eps)
\;\;.
\end{aligned} \right.
\end{equation}
Overall we have  $2 \times 2+2 \times 2+1+1+1= 11$ equations. 
In the EW framework, at difference with QCD, it is possible to express analitically
the initial conditions, since we are in a perturbative regime. At the $\mu=Q$ scale, 
these initial conditions can be expressed as:
\be 
f_{AB}(x,\eps=1)=\delta_{AB}\delta(1-x)
\;\;.
\ee

%%%%%%%%%%%%%%%%%%%%%%%%%%%%
\section{\label{sec3} Sum rules and cutoffs for the splitting functions}

In this section we show that sum rules on the splitting functions
enforce the presence of precisely determined cutoffs near $z=0$. 
Sum rules are requirements on the integrals over $z$ of $P_{ij}(z)$ and
$zP_{ij}(z)$, obtained from the conservation of the total momentum and of
the quantum numbers. 

In the literature, in all the works where IR corrections are resummed, \cite{Fadin:1999bq,Melles:2001ye,Ciafaloni:1999ub,Ciafaloni:2001mu}, 
 the cutoffs near $z=1$ have been considered.
%
%We remark that, in the literature, in all the works where IR corrections are resummed \cite{Fadin:1999bq,Melles:2001ye,Ciafaloni:1999ub}, the cutoffs near $z=1$
%appearing, for instance, in $P^R_{ff}(z)$ (see Eq.~\eqref{split1}), 
%have been related to both kinematical constraints and infrared dynamics \cite{Ciafaloni:2001mu}.
%
In the present work, we introduce for the first time, to the best of our knowledge,
additional cutoffs near $z=0$. 
%appearing in $P^R_{gg}(z)$ and $P^R_{gf}(z)$, 
%see eqs.~(\ref{split2})  and (\ref{split3}). 
We present in this Section  the sources of these new cutoffs, while their
quantitative relevance on the values of the distribution functions $f_{ij}(z,\eps)$ 
will be discussed in Sect.~\ref{sec4}.

Conserved quantities involve integrals of the distribution functions. 
For instance, if the initial particle is a neutrino, then 
the probability to become an electron is $\int dz f_{e \nu}(z,\eps) $. 
If we are interested in fermion number conservation, the initial fermion number, 1, must be conserved when we sum over all possible final states:
\ba
\nonumber
1 &=&\int_0^1 dz (f_{e \nu}(z,\eps)+f_{\nu\nu}(z,\eps)-f_{\bar{\nu}\nu}(z,\eps)-f_{\bar{e} \nu}(z,\eps))
\\ &=&
2\int_0^1 dz \ff{\lzm}{\nu}(z,\eps)=\int_0^1dz\ff{\lzm}{\lzm}(z,\eps)
\label{vwlll}
\;\;,
\ea
where we used Eqs.~(\ref{translateL0}-\ref{translateL2}) for both left and right indices, and 
considered the conservation of the weak isospin. This condition must be satisfied for every value of $\eps$.
This means that 
the evolution equations must have kernels which allows the conservation of the 
condition (\ref{vwlll}).

In order to explore the consequences of Eq.~(\ref{vwlll}), it is convenient to write the evolution 
in terms of the Mellin transform, i.e. by defining the $N$-th order moments:
\be\label{oiewoi76}
\mt(N,\eps)=\int_0^1 dz f(z,\eps) z^{N-1}
\;\;,
\ee
such that Eq. (\ref{sola}) become factorized, with the ordinary product:
\be\label{solamel}
\der \mt_{ij}(N,\eps)=\mt_{ik}(N,\eps) \,\mtp^G_{kj}(N,\eps)\qquad;\qquad
\der \mt_{AB}(N,\eps)=\mt_{AC}(N,\eps)\, \mtp^I_{CB}(N,\eps)
\;\;.
\ee
Using these equations in the $N=1$ case, and deriving Eq.~(\ref{vwlll}) with respect 
to $\eps$ we obtain:
\be
0=\der \Ff{\lzm}{\lzm}(1,\eps) =\frac{3}{4}\Ff{\lzm}{\lzm}(1,\eps)\,
(\mtp^R_{ff}(1,\eps)+\mtp^V_{ff}(1,\eps))
\;\;.
\ee
Since the value of $\Ff{\lzm}{\lzm}(1,\eps)$ depends on $\eps$, therefore it
 is arbitrary, the above equation is satisfied for every value of $\eps$ if 
\[
\mtp^R_{ff}(1,\eps)+
\mtp^V_{ff}(1,\eps)=0
\;\;.
\] 
We can now use Eq.~(\ref{split1}) to obtain:
\be\label{9y8}
\mtp^R_{ff}(1,\eps)+
\mtp^V_{ff}(1,\eps)=\int_0^1dz(P^R_{ff}(z,\eps)+P^V_{ff}(z,\eps))=2\eps+{\cal O}(\eps^2)\to 0
\;\;.
\ee
The sum rule is not exact, but leading terms in $\log \eps$ and subleading 
constant terms cancel with each other, 
while there is a remnant of (irrelevant) mass-suppressed terms. 

We consider now the momentum conservation. Since $z$ is the momentum fraction of the daughter (final) particle, considering a parent (initial) neutrino and 
operating as we have done in the case of the fermion number we have:
\be\label{oijew}
1=\sum_j\int_0^1 dz\, z f_{j\nu}(z,\eps)=2\Ff{\Lzp}{\nu}(2,\eps)+\sqrt{3}\Ff{\Gzp}{\nu}(2,\eps)=\Ff{\Lzp}{\Lzp}(2,\eps)+\frac{\sqrt{3}}{2}\Ff{\Gzp}{\Lzp}(2,\eps)
\;\;.
\ee
Analogously, for a neutral gauge boson initial state we obtain:
\be\label{lknjh}
1=\sum_j\int dz z f_{j W_3}(z,\eps)=2\Ff{\Lzp}{W_3}(2,\eps)+\sqrt{3}\Ff{\Gzp}{W_3}
(2,\eps)=\frac{2}{\sqrt{3}}\Ff{\Lzp}{\Gzp}(2,\eps)+\Ff{\Gzp}{\Gzp}(2,\eps)
\;\;.
\ee
We now use Eq.~(\ref{eqnonso}) in the $0^+$ channel:
\ban
&0=\displaystyle{\der}
\left(\Ff{\Lzp}{\Lzp}(2,\eps)+\frac{\sqrt{3}}{2}\Ff{\Gzp}{\Lzp}(2,\eps) \right)
\\
&=\frac{3}{4}\left[\left(\Ff{\Lzp}{\Lzp}(2,\eps)+\frac{\sqrt{3}}{2}\Ff{\Gzp}{\Lzp}(2,\eps)\right)
(\mtp^R_{ff}+\mtp^V_{ff})(2,\eps)+\left(\frac{2}{\sqrt{3}}\Ff{\Lzp}{\Gzp}(2,\eps)+\Ff{\Gzp}{\Gzp}(2,\eps)\right)\mtp_{gf}^R(2,\eps)\right]
\;\;,
\ean
and considering again Eqs.~(\ref{oijew} , \ref{lknjh}) we obtain the sum rule that can be seen to be obeyed by the splitting functions (\ref{split1}-\ref{split3}):
\be
\mtp^R_{ff}(2,\eps)+\mtp^V_{ff}(2,\eps)+\mtp^R_{gf}(2,\eps)={\cal O}(\eps)\rightarrow 0
\;\;.
\ee

Up to this point we have considered sum rules that the electroweak evolution equations have in common with those of 
the QCD. Indeed, in the case of strong interactions, due to confinement, initial states are color singlets, that 
correspond to the $T=0$  evolution equations. In the case of EW interactions instead, initial states have 
isospin quantum numbers and, therefore, further equations in the channels $T=1$ and $T=2$ are present. 
A consequence of this feature of the EW equations is that the  double logs terms of IR origin
do not cancel. This effect has been called ``Bloch-Nordsieck (BN) violation''  \cite{Ciafaloni:2000df}. 

We now proceed to show that EW evolution equations lead
to sum rules that are absent in QCD, and are responsible ultimately for the necessity of precise cutoffs in the splitting functions near $z=0$.

The conservation of the third component of the weak isospin, $T_3$, is related to the  $T=1,CP=-$ structure function. Taking a neutrino and a neutral gauge boson as initial particles,  we have:
\ba\label{consnu}
t^3_\nu&=&\frac{1}{2}=\sum_i t^3_i \mt_{i\nu}(1,\eps)
=\frac{1}{2}(\mt_{\Lum\Lum}(1,\eps)+\sqrt{2}\mt_{\Gum\Lum}(1,\eps))
\;\;,
\\
\label{consw}
t^3_{W^+}&=&1=\sum_i t^3_i \mt_{iW^+}(1,\eps)
=\frac{1}{\sqrt{2}}(\mt_{\Lum\Gum}(1,\eps)+\sqrt{2}\mt_{\Gum\Gum}(1,\eps))
\;\;.
\ea

We use the evolution equation (\ref{eq1m}) and the relations (\ref{consnu},\ref{consw}) 
and we operate as in the case of momentum conservation. 
We  finally obtain:
\ba\label{àpkàk}
&~&
3\mtp^{V}_{ff}(1,\eps)-\mtp^{R}_{ff}(1,\eps)+4 \mtp^{R}_{gf}(1,\eps)=0\Rightarrow
\mtp^{R}_{ff}(1,\eps)=\mtp^{R}_{gf}(1,\eps)
\;\;,
\\\label{orecchie}
&~&
\frac{1}{2}\mtp^{R}_{fg}(1,\eps)+ \mtp^{R}_{gg}(1,\eps)+2\mtp^{V}_{gg}(1,\eps)= 0
\;\;.
\ea
where the last step in Eq.~(\ref{àpkàk}) has been obtained by using Eq.~(\ref{9y8}).

Let us consider first the sum rule (\ref{àpkàk}). 
If we use the standard expression for $P^R_{gf}$, i.e. 
that without the cut close to $z=0$, i.e.
\be
P_{gf}^R(z)\to\frac{1+(1-z)^2}{z}
\;\;,
\ee
Eq.~(\ref{àpkàk}) cannot be satisfied since $\mtp^{R}_{ff}(1,\eps)$ is finite, while $\mtp^{R}_{gf}(1,\eps)$ diverges due to the singularity for $z\to 0$ proportional to $1/z$. 
If instead we impose a cutoff and use the expression for $\mtp^{R }_{gf}(1,\eps)$ 
of Eq.~(\ref{split3}), then the sum rule is satisfied. 

The following points are worth to be emphasized.
\begin{itemize}
\item
$\mtp^{R}_{ff}(1,\eps)$ is divergent for $z\to 1$ while $\mtp^{R }_{gf}(1,\eps)$ is divergent for $z\to 0$, therefore the sum rule connects the (known) cutoffs of IR origin close to 1 with the (new) cutoffs close to 0.
\item
The value of the cutoff $z>\eps$ is precisely determined by the requirement of 
Eq.~(\ref{àpkàk}). With a different value like $z>2\eps$ the sum rule would be violated.
\item
The presence of the cutoff is justified from a probabilistic point of view. 
We expect that the probability for a fermion to become a fermion of momentum fraction $z$ 
has to be the same of the probability of becoming a gauge boson of momentum fraction 
$1-z$: both are related to the same tree level diagram (see diagram $(a)$ in Fig.~\ref{splittings}).
\end{itemize}

Similar considerations hold for  $P^R_{gg}$, whose integral in $z$ must satisfy
Eq.~(\ref{orecchie}). This is obtained by introducing
the cutoff near $z=0$ which has been highlighted by the square brackets
in Eq.~(\ref{split2}). 

Finally, we point out
that the new splitting functions are $P_{gf}$ and $P_{gg}$ and they involve the splitting into a gauge boson. Indeed, the need for a cutoff close to 0 arises because of the
$1 / z$ singularity related to the IR dynamics of a soft gauge boson.

%%%%%%%%%%%%%%%%%%%%%%%%%%%%%%%%%%
\section{Quantitative results \label{sec4}}

In this section, we compare the PDFs obtained with the uncut splitting functions,
i.e. the standard ones without cutoff, with those obtained with our splitting 
functions which include the cutoff.

In order to illustrate the different features characterizing the two types of PDFs,
we consider first the analytical expressions of the first order terms of a perturbative expansion 
in powers of $\alpha$:
\be
f_{AB}(x,\eps)
=\sum_{i=0}^\infty \alpha^i f_{AB}^{(i)}(x,\eps)
\label{eq:expansion}
\;\;,
\ee
where the zeroth order term is dictated by the initial conditions
\be
f_{AB}^{(0)}(x,\eps)=\delta_{AB}\delta(1-x)
\;\;,
\ee
which indicate that, in the no-emission case, the probability 
of finding a particle inside itself is equal to the unity.

Using the perturbative expansion in Eqs.~(\ref{eqini}-\ref{eq1m}), 
we obtain analytical results for all first order terms of the PDFs. 
For instance, in the case of the $(\Lzp , \Gzp)$ channel we have
\be\label{asfd}
f_{\Lzp,\Gzp}^{(1)}(x,\eps)= \frac{1}{\pi}\frac{\sqrt{3}}{2}
\int_\eps^1\frac{d\eps'}{\eps'}P^R_{fg}(x,\eps')
\;\;,
\ee
and we obtain two different results for our, cut (c), and 
standard, uncut (u), cases:
\be
\ff{\Gzp}{\Lzp}^{(1)c}(x,\eps)=\frac{1}{\pi}\frac{\sqrt{3}}{2}\frac{1+(1-x)^2}{x}
\theta(x-\eps)\log\frac{x}{\eps}\qquad ; \qquad
\ff{\Gzp}{\Lzp}^{(1)u}(x,\eps)=\frac{1}{\pi}\frac{\sqrt{3}}{2}\frac{1+(1-x)^2}{x}\log\frac{1}{\eps}
\;\;.
\ee

Already from these first order expressions we see that the differences between the PDFs obtained 
with our and standard splitting functions are relevant. Indeed, in 
the standard case the behaviour close to 0 is $\propto 1/x$ and therefore divergent, 
while in our case the PDF is $\propto (1 /x) \log(x/\eps)$ therefore it is
continuous and, because of the step function, is 0  for $0 \le x \le \eps$. 

In addition to the first order, we obtained the complete solutions 
by numerically solving  Eqs.~(\ref{eqini}-\ref{eq1m}). 
We consider  their evolution in the variable
 $t\equiv\log\eps=\log(\mu/Q)$,  which ranges in the interval $[t_{\min},0]$ with
 $t_{\min}\equiv\log(M/Q)<0$. 
 We used a discrete two-dimensional grid $[t_{\min},0]\times[0,1]$ in the
 $(t,x)$-plane to transform the differential equation~\eqref{sola} into a finite-difference one. 
 The $z$-integration occurring in Eq.~\eqref{convolution} has been carried out
 by using an adaptive
 method which evaluates the exact values of the splitting functions and an
 interpolation of the evolving PDFs. This procedure has the advantage of a
 better treatment of the $t$-dependent cuts on the $z$-integration, since the
 boundaries of the $z$-integration are almost never found on the discretized
 $x_i$ points. We implemented the $t$-evolution via a 4-point Runge-Kutta
 algorithm and increase the number of points in the grid until we reach the
 required  precision.
 
We studied the impact of the new constraints on the splitting functions by solving
Eqs.~(\ref{eqini}-\ref{eq1m}) with and without the $\theta(x-\eps)$ terms 
in the expressions (\ref{split2}) and (\ref{split3}) of $P^R_{gg}$ and $P^R_{gf}$. 
We carried out calculations for three different values of $\eps$, specifically 
0.01, 0.001, 0.0001.
%$10^{-2}$, $10^{-3}$, $10^{-4}$.
We present here only some selected results for $\eps=0.01$, since the physics 
contents of the other cases is  analogous. 

\begin{figure}[htb]
   \centering
    \includegraphics[scale=0.6,angle=90] {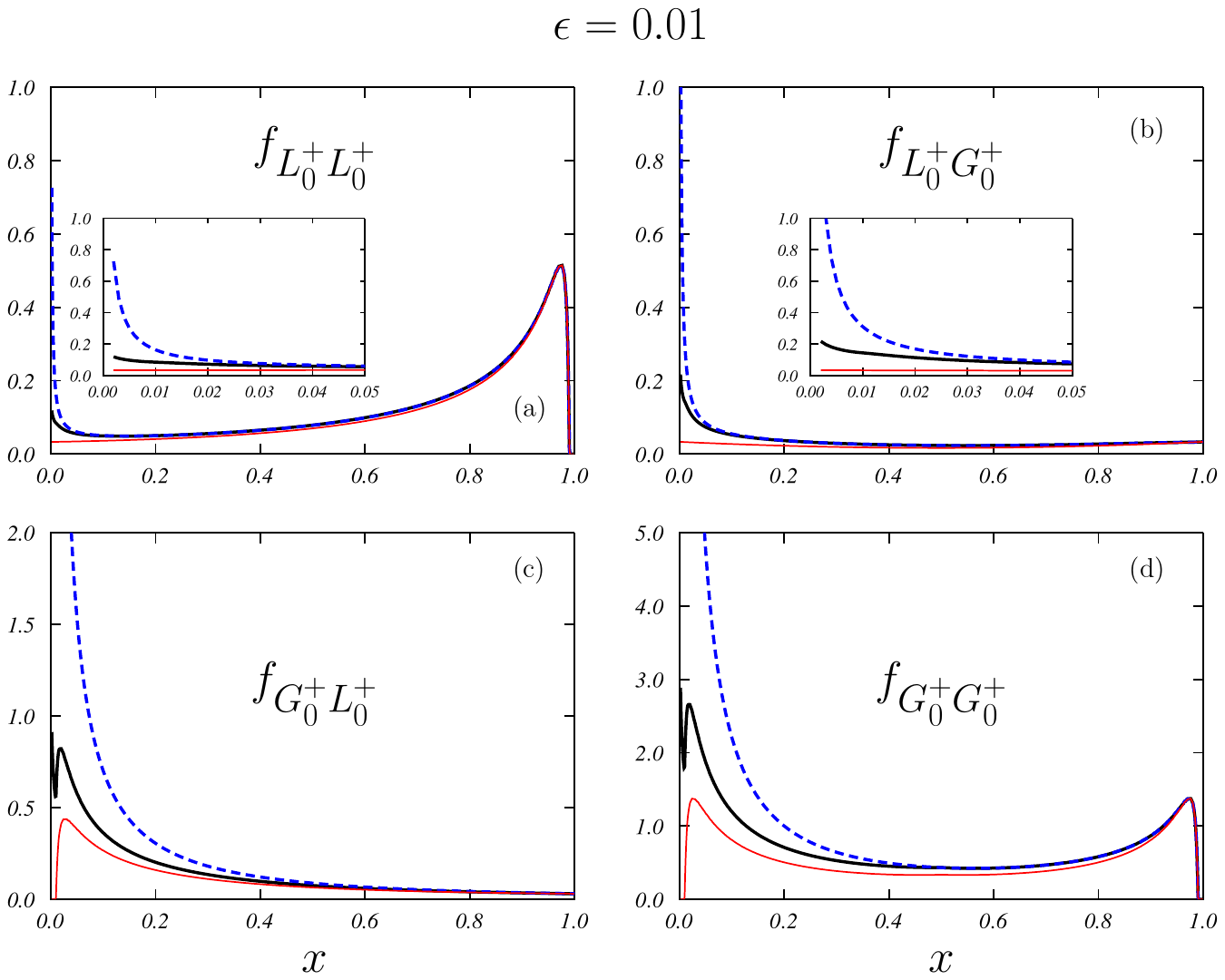}
    \caption{
 PDFs of the $T=0$ isospin channels. The full black lines show the results of our full calculations 
 where the new constraints on the $P^R_{gg}$ and $P^R_{gf}$ splitting functions have been considered. 
 The thin red lines show the results obtained by considering only the first order term in the $\alpha$
 power expansion of Eq.~(\ref{eq:expansion}). The results of the complete, numerical, calculation
carried out without considering the constrains in the splitting functions are shown by the blue dashed
lines. 
    }  
\label{fig:ris1}  
\end{figure}

We show in Fig.~\ref{fig:ris1}  the PDFs as a function of the momentum fraction
$x$ for the four $T=0$ isospin channels. The PDFs obtained with our
splitting functions, $f^c$, are indicated by the black lines,
while those obtained with the standard splitting functions, $f^u$, by the dashed 
blue lines. The thin red lines show the first order solution $f^{(1)c}$
in an expansion in powers of $\alpha$ (see Eq.~\ref{eq:expansion}).
%%%%%%%%%%%%%%%%%%%%%%%%%%%%%%%%%%%%%%
% Canali con isospin
%%%%%%%%%%%%%%%%%%%%%%%%%%%%%%%%%%%%%%

As expected, the differences between $f^c$ and $f^u$ are more significant at 
small values of $x$. We emphasize these difference in the insets of the 
panels (a) and (b) where we show the PDFs for  $x \ll 1$.

\begin{figure}[htb]
      \centering
      \includegraphics[width=80mm]{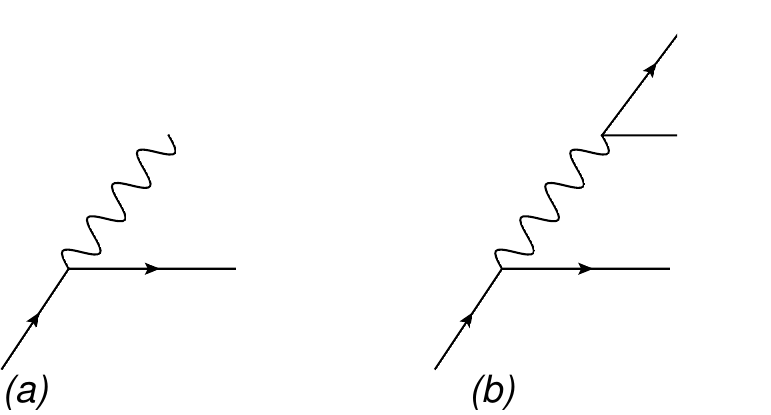}
     \caption{ \label{splittings} First order contribution to $f_{GL}$ (a) and second order contribution to $f_{LL}$ (b) }  
      \end{figure}

 These results show that the difference between $f^c$ and $f^u$ is significant in the isospin channels describing gauge bosons distributions like $\ff{\Gzp}{\Gzp}$ and $\ff{\Gzp}{\Lzp}$, while in the channels describing lepton distributions like $\ff{\Lzp}{\Gzp}$ and $\ff{\Lzp}{\Lzp}$ these differences are less pronounced. 
The reason for this is that the new constraint are present in $P_{gf}$ and $P_{gg}$ and they involve the splitting into a gauge boson. These splitting functions contribute at first order for a gauge boson distribution but only at second order for a lepton distribution (see Fig. \ref{splittings}). For this reason the effect of the new constraints is more pronounced for gauge bosons distributions.

%%%%%%%%%%%%%%%%%%%%%%%%%%%%%%%%%%%%%%
% Diagramma primo e secondo ordine
%%%%%%%%%%%%%%%%%%%%%%%%%%%%%%%%%%%%%%			

We point out that the effect of the new constraints is not limited to very small values of $x$: for example, for the complete solution, in the case of
 $\ff{\Gzp}{\Lzp}$, at $x=0.4$ $f^u$ is bigger than $f^c$ by 25\%.
This could surprise because our splitting functions differs significantly from the standard ones
only in the region $x \simeq \eps$, and we choose very small values of $\eps$. 
However, one has to consider that in the EW evolution  equations, the splitting functions are
integrated on all the possible values of $\eps'$, i.e. from $\eps'=\eps$ to $\eps'=1$,
and this means that the new cutoffs generate effects for every value of $x$.

\begin{figure}[htb]
      \centering
      \includegraphics[scale=0.6,angle=90] {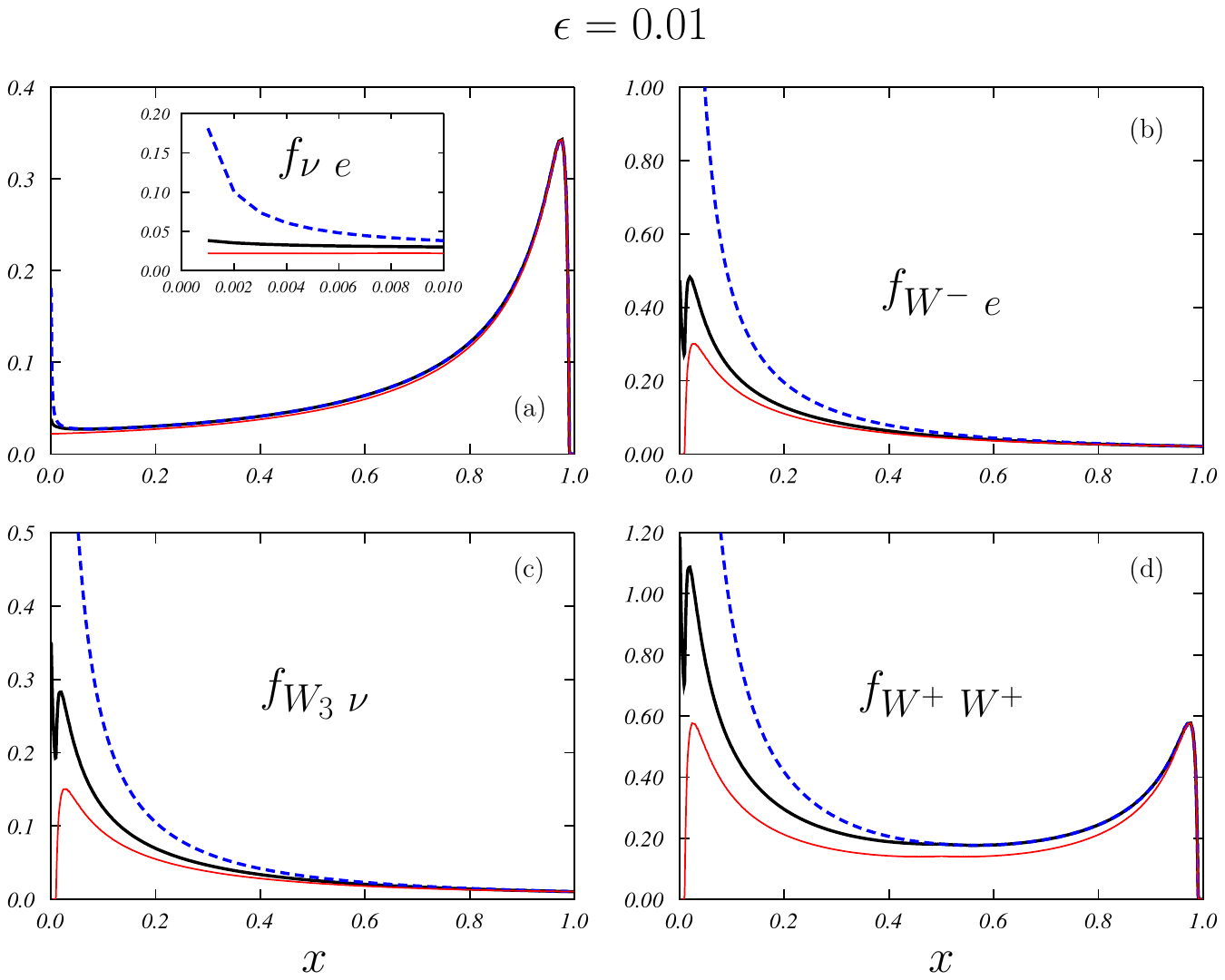}
     \caption{ \label{fisiche}   
     PDFs in four selected physical channels. The meaning of the lines is the same 
     as in Fig.~\ref{fig:ris1}. 
     }  
\end{figure}

We have evaluated the PDFs in the physical channels by inverting the $U$ matrix of Eq.~(\ref{transformer}). We show in Fig.~\ref{fisiche} the results for four selected channels
representative of the 49 ones. The meaning of the lines is the same as in Fig.~\ref{fig:ris1}. 
%%%%%%%%%%%%%%%%%%%%%%%%%%%%%%%%%%%%%%
% Canali fisici
%%%%%%%%%%%%%%%%%%%%%%%%%%%%%%%%%%%%%%			

In the isospin representation we have pointed out that the gauge boson channels are 
most sensitive to the new cutoffs, and this is reflected also in the physical channels.

The results of the panel (a) of Fig.~\ref{fisiche} show that the new constraints modify 
the PDFs only at very small values of $x$, as it is shown in the inset. This is a purely leptonic
channel with an electron producing a neutrino. In this case, the three curves almost 
overlap, except when $x\simeq 0$. 
%Even the solution of the first order expansion 
%term well describes the behaviour of the complete solution. 

The situation is very different when, at least, one gauge boson channel appears, as it is shown
in the other panels of the figure. The relative difference between $f^c$ and $f^u$ at $x=0.4$
is about 22\% in the $(W^-,e^-)$ and $(W_3,\nu)$ channels, panels (b) and (c), 
and about 10\% in the $W^+,W^+$ channels, panel (d).

In these channels the solution for the first order expansion term fails in describing the correct
behaviour of the complete solution. The difference in the $W^+,W^+$ channel is remarkable
even from the qualitative point of view. The first order solution is symmetric around $x=0.5$,
while the full solution is clearly asymmetric with values at small $x$ remarkably larger than 
those around $x=1$. 

%%%%%%%%%%%%%%%%%%%%%%%%%%%%%%%%%%%%%%%%%%%%%%

\section{\label{sec5}IR and UV evolution equations}

The EW evolution equations were originally derived as IR equations \cite{Ciafaloni:2001mu,Ciafaloni:2005fm}, i.e. equations where the varying scale is an infrared parameter $\mu$ having the meaning of lower bound on transverse momentum of the emitted particles. From an alternative point of view,
different groups \cite{Bauer:2017isx,Garosi:2023bvq} used
UV evolution equations, where the varying scale is a parameter $q$ having the meaning of upper bound on the transverse momentum of the emitted particles. It is not clear that the two approaches produce the same results. 
In this section, we show that the solutions of the UV and IR evolution equations coincide,
when the same splitting functions, including the cutoffs, are consistently used in both approaches. 
%%%%%%%%%%%%%%%%%%%%%%%%%%%%%%%%%%%%%%
% Diagrammi evoluzione
%%%%%%%%%%%%%%%%%%%%%%%%%%%%%%%%%%%%%%			
\begin{figure}[htb]
      \centering
      \includegraphics[width=150mm]
                  {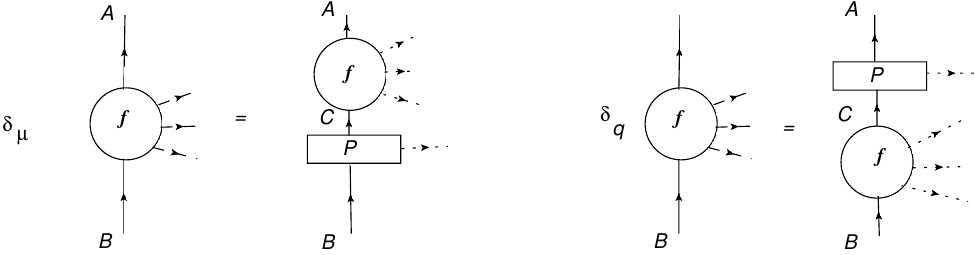}
     
 \caption{\label{IRUV}   IR (left) and UV (right) evolution equations. 
 The dashed lines represent particles belonging to the final state.}  
      \end{figure}

We consider splitting functions depending not only on the momentum fraction
 $z$ but also on the transverse momentum $\kt$: $P_{AB}(z,\kt)$. In the IR equation the contribution of the emitted particle with lower momentum is factorized (fig. \ref{IRUV},left side); calling $\mu$ the lower cutoff on this momentum we have:
\be\label{eqir}
\mu\frac{\de}{\de \mu} f^{IR}_{AB}(x;\mu,Q)=-f_{AC}^{IR}\otimes P_{CB}(\mu)
\qquad ; \qquad
f^{IR}(x,\mu=Q)=f_0(x)
\;\;.
\ee
In the case of UV equations instead, the contribution of the hardest emitted particle is factorized (fig. \ref{IRUV},right side). 
We call $q$ the upper bound on the transverse momentum and obtain:
\be\label{equv}
q\frac{\de}{\de q} f^{UV}_{AB}(x;M,q)=P_{AC}(q)\otimes f^{UV}_{CB}  
\qquad ; \qquad
f^{UV}(x,q=M)=f_0(x)
\;\;.
\ee
The formal solution, in the IR case, is given by:
\be\label{solir}
f^{IR}(\mu,Q)=f_0\otimes {\rm K}\{\exp[P]\}(\mu,Q)
\;\;,
\ee
having defined:
\ba
\nonumber
{\rm K}\{\exp[P]\}_{ij}(\mu,q)&=&\delta(1-z) \delta_{ij} +\int_\mu^{q} P_{ij}(k)\frac{dk}{k}
\\ 
&+&\sum_{n\ge 2}\int\displaylimits_{\mu<k_n<k_{n-1}<\cdots<k_1<q}(\Pi_i\frac{dk_i}{k_i})P_{ii_1}(k_1)
\otimes P_{i_1i_2}(k_2)\otimes\cdots
\otimes P_{i_nj}(k_n)
\;\;.
\label{ame}
\ea
Eq.~(\ref{solir})  is indeed solution of Eq.~(\ref{eqir}), since 
the only variable depending on $\mu$ in the ordered product of Eq.~(\ref{ame})  is $k_n$. 
When we derive with respect to $\mu$ we obtain (minus) the value of the integrand calculated for $k_n=\mu$ and this gives the solution of (\ref{eqir}). It is then sufficient to 
implement the initial condition $f_0$. 

A similar reasoning applies to the UV case. 
In this case, only $k_1$ depends on $q$ and we have:
\be\label{soluv}
f^{UV}(M,q)= {\rm K}\{\exp[P]\}(M,q)\otimes f_0
\;\;.
\ee
Since we eventually set $\mu= M$ in (\ref{solir}) and $q=Q$ in (\ref{soluv}), the IR case  and UV case  differ only because of a different placement of the $f_0$ function. However, since we have
$f_0=\delta(1-z)\delta_{AB}$ and since $\delta(1-z)$ is the identity for convolutions and $\delta_{AB}$ is the identity in the $P$ matrices space, the two solutions are equal:

\be
f^{IR}(M,Q)=f^{UV}(M,Q)= {\rm K}\{\exp[P]\}(M,Q)
\;\;.
\ee
This establishes the equivalence between the IR and the UV approaches.

 As a final comment, we remark
 that Eqs.~(\ref{eqir}) and (\ref{equv}) are equivalent, i.e. they produce the same solutions for the PDFs, if, and only if, the same splitting functions are used, albeit with different arguments. 
 If, as in the case treated here, 
 there is a cutoff for $x<1- \mu / Q$ in $P_{AB}(z,\mu)$ for the IR equations, 
 then, in the ultraviolet equations a cutoff at $x<1- q / Q$ must appear in $P_{AB}(z,q)$. 
 This might sound trivial, but the point is that in the literature different cutoffs appear.
 For instance in \cite{Bauer:2017isx}
the cutoff is rather $z<1- M/q$, leading, in general, to PDFs that differ from 
those we obtain. 

%%%%%%%%%%%%%%%%%%%%%%
\section{\label{sec:conclusions} Summary, conclusions and perspectives}

At c.m. energies much grater than   the weak scale, energy growing Electroweak Radiative corrections can be taken into account by defining Parton Distribution Functions (PDFs) that obey Electroweak Evolution Equations (EWEE), in analogy with DGLAP in QCD.
  In this work we propose to modify EWEEs with respect to what has been done until now in the literature and we analyze the impact on PDFs of these modifications. 
In particular, electroweak interactions  are characterized by isospin 1 evolution equations
that are absent in the corresponding isospin 0  QCD (DGLAP) and QED equations.
Isospin conservation, related to these isospin 1 equations, requires to modify the splitting functions (that are the kernels of EWEEs) by adding suitable cutoffs.
The solutions (PDFs) obtained with these new kernels differ significantly from the ones using the standard kernels used in the literature until now (see fig. \ref{fisiche}).

In this work we have also addressed the issue of comparing the results obtained by using a IR approach (ours) with previous results obtained using a UV approach, as is customary in the literature. We have shown that UV equations indeed produce the same PDFs as IR equations, but only if a careful choice of the cutoffs in the splitting functions is made. Finally, let us note that work has still to be done in order to provide theoretical results that can be compared with the experimental measurements. First, QCD and QED interactions have to be added and then the full particle spectrum of the Standard Model has to be considered, while we chose to consider only a subset. Moreover, for small momentum fraction $x$, additional terms proportional to $\log x$ will possibly have to be added to the equations \cite{Ciafaloni:2008cr}.

 The modifications described here will be particularly relevant if
a 100 TeV hadronic collider and/or a TeV scale muon collider will see the light.
%%%%%%%%%%%%%%%
\vskip .8 cm 
AKNOLEGDMENTS. 
This project has been partially supported by the European Union’s Horizon
2020 research and innovation programme under grant agreement N$^\circ$ 824093.

%%%%%%%%%%%%%%%%%%%%%%%%%%%%%%%%%%%%%%

\end{document}